# Topological and Semantic Graph-based Author Disambiguation on DBLP Data in Neo4j

Valentina Franzoni§†, Michele Lepri°, Alfredo Milani†

§ Department of Computer, Control, and Management Engineering, Sapienza University of Rome, Italy
† Department of Mathematics and Computer Science, University of Perugia, Italy
valentina.franzoni@dmi.unipg.it ✉
° Engineers Professional Association of Perugia, Italy

*Abstract*—In this work, we introduce a novel method for entity resolution author disambiguation in bibliographic networks. Such a method is based on a 2-steps network traversal using topological similarity measures for rating candidate nodes. Topological similarity is widely used in the Link Prediction application domain to assess the likelihood of an unknown link. A similarity function can be a good approximation for equality, therefore can be used to disambiguate, basing on the hypothesis that authors with many common co-authors are similar. Our method has experimented on a graph-based representation of the public DBLP Computer Science database. The results obtained are extremely encouraging regarding Precision, Accuracy, and Specificity. Further good aspects are the locality of the method for disambiguation assessment which avoids the need to know the global network, and the exploitation of only a few data, e.g. author name and paper title (i.e., co-authorship data).

*Keywords—disambiguation; entity resolution; semantic networks; topological similarity; link prediction; bibliometrics; graph-based database; Neo4j; NoSQL Database; co-authorship network; social network analysis*

## I. INTRODUCTION

The diffusion of online social networks with ubiquitous access and redundant data can be a challenge for using additional information in order to prevent ambiguity. *Entity Resolution (ER* is the task of identifying and linking different manifestations of real-world objects, compensating for the lack of data quality, such as incorrect or missing data or attributes change over time. ER is essential for several applications, starting with simple de-duplication tasks. For instance, for a company or for a political party, which rely on a database of people who declared their interest in a product/service/activity or were automatically added to the list, to promote or send an advertisement to a de-duplicated set of people could be an economic advantage. More complex applications include, among others, recommender systems, social network analysis, and digital library management for *research bibliometrics*. In the latter, authors and other inventory metadata should be unambiguously identified and referred to the organisation/topic/journal context related to their production.

*Disambiguation* is a solution to the Entity Resolution problem, recognising and clustering such different object manifestations in groups corresponding to the same object, i.e. identifying which representation objects are different phenotypes of the same real object. Disambiguation is, in fact, the task of identifying entities referring to the same object in a set of candidates [1], [2].

*Link Prediction (LP)* [3] is a method establishing the strength of ties in a graph, e.g. a social network [4] or a bibliographic database. The more the strength, the more robust the ties among potential collaborators, e.g. co-authors.

*Topological similarity* is a suitable candidate method for link prediction. We hypothesise that authors with many *common co-authors* are similar [5], [6]. In fact, if anyone looks at himself as if he was another person, it is easy to guess that a proper evaluation function for similarity will give a high evaluation to such two ambiguous candidates (i.e., him as an observer and him as observed).

In this project we propose a new approach for link prediction using topological (e.g. Adamic, Common Neighbours) and semantic similarity (e.g. Confidence, PMI and PMING). We aim to predict future links or infer existing unknown links from a social network on Neo4j as a graph-based database. The workflow for the experimentation, performed on DBLP, includes the following phases: import DBLP in Neo4j, import information about links (e.g. authorship, publication), import useful metadata (e.g. year of publication), link prediction through graph-based data query, result evaluation. The main idea is that it is useful for link prediction to use a graph-based database in order to quickly perform queries on the links on common-neighbours, in a much more efficient way than on SQL-based databases, or even edges lists, especially for long paths.

Furthermore, our experiments apply, to bibliometric analysis, previous models for semantic similarity and link prediction, stating that it is possible to generalize semantic similarity under the assumption that similar concepts co-occur in documents/databases indexed by a search engine, and that the information on common-neighbours of common-neighbours (up to a maximum path length of three) are effective and efficient for link prediction. Both semantic and topological similarity measures make profit from suboptimal results of computations, where approximations to evaluate frequencies and probabilities can be used to calculate semantic proximity or distance. Which measure to use, and how to optimize the extraction and the utility of the extracted information, are open issues to which we contribute with our results.

In the following sections, we will define and explain the Entity Resolution problems of data quality, consider topological similarity for link prediction as a potential disambiguation method and expose our experiments in the realm of bibliographic data using DBLP [7] in Neo4j [8]. We will use a 2-steps traversal disambiguation with topological-similarity-based link prediction, then discuss the results of our preliminary experimentation phase and present conclusions about the proposed approach.

## II. DATA QUALITY AND DISAMBIGUATION

Widely recognised properties for data quality include the following [9]:
- *Relevance*. Data meet the needs for which they are collected.
- *Accuracy*. Data cannot be protected against every error, but the grade of accuracy of the primary variables of interest can be defined and evaluated.
- *Timeliness*. Concerns about how current does the information need to be, in order to make predictions.
- *Comparability*. Defines if it is appropriate to compare several databases to facilitate data use in modeling statistical estimators.
- *Coherence*. Data are logically interconnected and consistent.
- *Completeness*. No records are missing; no records have missing elements.

Unfortunately, many organisational data do not meet these requisites. In the case of bibliometric research data, poor data quality can be due to situations such as:
- *Incorrect data:* values are recorded incorrectly in the data entry phase, e.g. typographical or spelling mistakes, or information is incorrect, e.g. the wrong postal code in a postal address.
- *Missing data*: values are not recorded, e.g. missing university department and address information, when only the name of the academy is given.
- *Attributes change* over time, e.g. an author changing affiliation, or an institute changing the name, as happened in last five years in Italian universities, where faculties were eliminated as institution level, and departments changed names and sometimes merged.
- *Irrelevant data:* relevant information is submerged in poorly structured data, e.g. poor or diverse data formatting.
- *Imprecise data:* information is not recorded at the desired level of granularity, making a wrong use of a proper format, e.g. name and surname inverted, wrong level on the specificity of the affiliation (for example confusing laboratory or research group, a department with faculty or university).

Furthermore, acronyms, abbreviations, and data truncation, among others for space limitations, can vary from publication to publication. Translations can also introduce issues, e.g. original name versus English name, or different translations of the same institutional name. Although most authors indicate their affiliation with an English name, translation from the original entity name may not be univocal, and therefore difficulties may arise to credit the same institution for publications in which it was indicated with different names.

An excellent approach to the solution will include prevention, detection and repair of disambiguation [10]–[12]. Henk Moed [13] assesses that the best disambiguation method is the human manual one.

## III. TOPOLOGICAL SIMILARITY FOR LINK PREDICTION

Common Neighbours-based (CN) rankings (e.g. Jaccard, Adamic-Adar) [14], [15] represent a class of similarity measures for link prediction, which efficiently assess the likelihood of a new link based on the neighbours frontier of the already existing nodes [16]. The CN similarity measures are more performant than others (e.g. preferential attachment, path-based distances [17]–[21]) in link prediction [22], [23], but they present the drawback of returning a large *0-tail*: a zero-rank value is given to all the links of pair of nodes, which have no common neighbours. Even if the rating is equal to *0*, such links may be potentially suitable for link prediction [24], [25].

Let a training network $G=(V,A)$ at time $t$, with $N$ nodes and undirected arcs $A \subseteq V x V$, and let a given test network $G'=(V,A')$, where $G'$ is an extension of $G$ at time $t+k$ and $A \subseteq A'$. The classic link prediction problem is to rank with a function $r(G)$ the set of potential links $A_{pot}$ which could appear in *G'*, such that the new links $A_{new}$ are ranked first. A perfect rank function (Kendall 1938) would return all the $A_{new}$ links in the first $|A_{new}|$ ranking positions:

$$KtA = \frac{1}{|A_{pot}|-1-|A_{new}|} * \sum_{l=|A_{new}|}^{|A_{pot}|-1} (r_l - |A_{new}| + 1) \quad (1)$$

We provide here the definition of three common-neighbours-based similarity measures $\Phi$: the simple Common Neighbours (CN), the Adamic-Adar (AA), and the Pointwise Mutual Information (PMI) [26].

Given two papers *p1* and *p2* of authors *a1* and *a2*:

- $\Phi_1 = CN(a1,a2)$ is a simple query in *Cypher*, the Neo4j query language, which counts the number of elements in the path like the following:
  ```
  match
  p=(a1:Author)-[*..2]-(a2:Author)
  where   a1.AuthorName="a1.name"   and
  a2.AuthorName="a2.name"
  return p as path;                    (2)
  ```
- $\Phi_2 = PMI(j1,j2) = log[p(j1,j2)(p(j1)p(j2)]$ (3)
- $\Phi_3 = AA(a1,a2) = Sum(1/log[degCN(a1,a2)])$ (4)

Note that (3) is originally a semantic similarity measure [27], [28] of which we provided a novel topological definition in a graph.

## IV. TWO-STEPS TRAVERSAL DISAMBIGUATION

Assume to start from a bipartite graph *authors-papers-authors* (see Fig.1), where authors may be ambiguous or partially disambiguated: this is the typical situation of a bibliographic database like the public DBLP Computer Science database when a new article is indexed. There can be specific information indicating an author identifier or, more likely, such information is not present. In particular, DBLP is extremely poor of information about the content of the papers.

The hypothesis is to have two authors *a1* and *a2* of two different papers *p1* and *p2*: we aim to evaluate through each paper the similarity of the two authors, i.e. the likelihood that *a1* and *a2* are the same person (i.e., *a1==a2*). This problem is present only if *a1* and *a2* have similar names, or the same name or abbreviation, i.e. *a1.fullName==a2.fullName* or *a1.abbreviatedName/Surname==a2.abbreviatedName/Surname*.

The two-steps traversal consists of 3 phases:

- *Phase 1:* starting from the node *a1*, evaluate her co-authors $a1_1...a1_n$ for her paper *p1*.
- *Phase 2:* from the node *a2*, evaluate her co-authors $a2_1...a2_n$ for her paper *p2*.
- *Phase 3:* the similarity measure $\Phi$ is applied to *p1* and *p2*, i.e. to the occurrences of authors and co-authors of *p1* and *p2*.

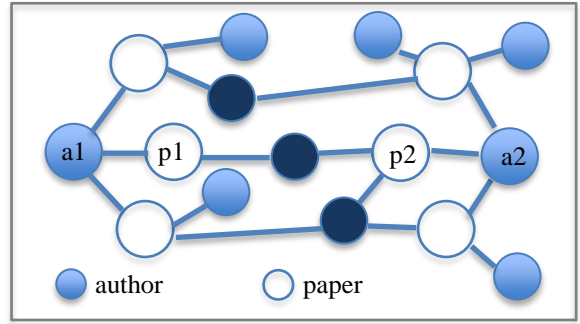

*Fig. 1: Two-steps disambiguation: two-steps(a1,a2)= $\Phi$(p1,p2), where p1 is a paper of a1 and p2 is a paper of a2, and $\Phi$ is a similarity measure*

Please note that, for the nature of the similarity measures presented in section III, paper nodes without common neighbours will have similarity equal to *0*, for the phenomenon of the *0-tail* [24].

Moreover, concerning classical Link Prediction (LP), the similarity of the two ambiguous candidates *a1* and *a2* is not directly calculated with $\Phi(a1,a2)$, but applying the measure to the papers of *a1* e *a2*. The two-steps traversal consists in navigating the path *author-paper-author* two times.

It is noticeable that the two-steps evaluation requires only local visits to the graph between *a1* and *a2*, being very efficient.

All the $\Phi_i$ topological similarity measures for LP have been slightly modified so that the function excludes the initial authors from the neighbour count of *p1* and *p2*.

For our experiments, we chose the graph-based database *Neo4j* as a proper NoSQL database for quick path traversal queries. The main program has been developed in *C#* using Neo4j APIs. As a dataset, a dump of the online Computer Science Database DBLP has been imported in Neo4j.

As an ambiguous author name, we chose to use the Chinese name *Chen LI*. LI is among the most frequent surnames in China, and Chen Li is among the most ambiguous names in DBLP, present with 24 versions, 23 of which are disambiguated with an identifier (see Fig.2), and 1 is a big furball of ambiguous LIs. The total amount of papers written by all the Chen LIs in DBLP is 72295.

- Chen Li
- Chen Li 0001
  University of California, Irvine, CA, USA
- Chen Li 0002
  South China University of Technology, Guangzhou, China
- Chen Li 0003
  New York University, Courant Inst. of Mathematical Sciences
- Chen Li 0004
  Purdue University
- Chen Li 0005
  Beijing Information Technology Institute, Beijing, China
- Chen Li 0006

Fig. 2: A subset of the Chen LIs in DBLP

The disambiguation classification of DBLP has been used as a Ground Truth, while our algorithm does not include any knowledge about the identifiers of the Chen LIs.

Noticeable that the database also presents a particular case of a paper co-authored by two different Chen LIs, one of whom is ambiguous and the other disambiguated, providing a further detail in the Ground Truth.

In our experiments, all the pairs of papers written by Chen LI have been evaluated, producing a similarity assessment for each pair of candidate Chen LIs. The similarity measures $\Phi_i$ applied are $\Phi_1$=*Common Neighbours (CN)*, $\Phi_3$=*Adamic-Adar (AA)*, and $\Phi_2$=*Pointwise Mutual Information (PMI)*, which are among the most performant for Link Prediction.

A rating of the pairs of authors have been obtained, where the value of $\Phi$ is considered positive (i.e. *a1==a2*) if greater than a threshold $\rho$. In the baseline experiment, $\rho$=0. Comparing the resulting classification with the DBLP Ground Truth (i.e. disambiguated Chen LIs with identifiers), we verified the classes of *True Positive (TP)*, *False Positive (FP)*, *True Negative (TN)*, and *False Negative (FN)* and from the count of such classes, we calculated the evaluation metrics of *Precision*, *Accuracy*, *Specificity* and *Sensitivity*.

## II. Discussion of Results

Experimental results are very good for Precision, Accuracy and Sensitivity. The experimented similarity measures all have good results: AA and CN have excellent performances, while PMI shown to be less performant, still obtaining Precision around 70% and Accuracy around 80%.

It is noticeable that the negative rate includes the 0-tail, for the cases where two papers of the same author do not have common co-authors: in such cases, the value of the similarity measures based on common neighbours is equal to *0*, therefore affecting the false negative classification. The 0-tail is particularly visible in the evaluation of Sensitivity, i.e. Recall, i.e. True Positive rate=*TP/(TP+FN)*, where the FN classification influences the total value. On the other hand, the positive classification is exact and performant (Specificity, i.e. True Negative rate=*TN/(TN+FP)* is greater than 0.95% for all the similarity measures), because of a low FP count. The only errors are on the papers in common for the Chen LIs who co-authored the same paper.

The result of the direct application of our approach leads to partitioning the ambiguous set of candidate authors in clusters of common authors. In order to avoid such partitions, a solution may be to apply a transitive closure of the relation *"is the same author"* obtained by the disambiguation method or to use a significative third category of data (i.e. metadata, e.g., affiliation, publication, year) to reduce the false negatives.

## VI. Conclusions

We introduced a novel local method in bibliometrics for authors disambiguation, applying topological similarity measures on a graph-based database in its implementation on DBLP data on Neo4j. Among the definition of the used topological similarity measures, we included a new topological definition of the Pointwise Mutual Information, initially a semantic similarity measure.

Experiments have shown very encouraging results: a deepening study is in course.

## References

[1] A. A. Ferreira, M. A. Gonçalves, and A. H. F. Laender, "A brief survey of automatic methods for author name disambiguation," *ACM SIGMOD Rec.*, vol. 41, no. 2, p. 15, 2012.

[2] A. Strotmann and D. Zhao, "Author name disambiguation: What difference does it make in author-based citation analysis?," *J. Am. Soc. Inf. Sci. Technol.*, vol. 63, no. 9, pp. 1820–1833, 2012.

[3] D. Liben-Nowell and J. Kleinberg, "The Link Prediction Problem for Social Networks," *Proc. Twelfth Annu. ACM Int. Conf. Inf. Knowl. Manag.*, no. November 2003, pp. 556–559, 2003.

[4] A. Milani and V. Franzoni, "Soft behaviour modelling of user communities," *J. Theor. Appl. Inf. Technol.*, vol. 96, no. 1, 2018.

[5] V. Franzoni, Y. Li, and P. Mengoni, "A Path-based Model for Emotion Abstraction on Facebook Using Sentiment Analysis and Taxonomy Knowledge," in *Proceedings of the International Conference on Web Intelligence*, 2017, pp. 947–952.

[6] V. Franzoni, G. Biondi, and A. Milani, "A web-based system for emotion vector extraction," in *Lecture Notes in Computer Science (including subseries Lecture Notes in Artificial Intelligence and Lecture Notes in Bioinformatics)*, 2017, vol. 10406 LNCS, pp. 653–668.

[7] M. Ley, "The DBLP computer science bibliography: Evolution, research issues, perspectives," in *Lecture Notes in Computer Science (including subseries Lecture Notes in Artificial Intelligence and Lecture Notes in Bioinformatics)*, 2002.

[8]   D. Dominguez-Sal, P. Urbón-Bayes, A. Giménez-Vañó, S. Gómez-Villamor, N. Martínez-Bazán, and J. L. Larriba-Pey, "Survey of graph database performance on the HPC scalable graph analysis benchmark," in *Lecture Notes in Computer Science (including subseries Lecture Notes in Artificial Intelligence and Lecture Notes in Bioinformatics)*, 2010.

[9]   J. Han, M. Kamber, and J. Pei, *Data Mining: Concepts and Techniques*. 2012.

[10]  G. Papadakis, G. Koutrika, T. Palpanas, and W. Nejdl, "Meta-blocking: Taking entity resolution to the next level," *IEEE Trans. Knowl. Data Eng.*, 2014.

[11]  A. Strotmann and D. Zhao, "Author name disambiguation: What difference does it make in author-based citation analysis?," *J. Am. Soc. Inf. Sci. Technol.*, 2012.

[12]  D. R. Amancio, O. N. Oliveira, and L. D. F. Costa, "On the use of topological features and hierarchical characterization for disambiguating names in collaborative networks," *EPL*, 2012.

[13]  H. F. Moed, "Measuring contextual citation impact of scientific journals," *J. Informetr.*, 2010.

[14]  J. Leskovec, D. Huttenlocher, and J. Kleinberg, "Predicting Positive and Negative Links," *Int. World Wide Web Conf.*, pp. 641–650, 2010.

[15]  L. A. Adamic and E. Adar, "Friends and neighbors on the Web," *Soc. Networks*, vol. 25, no. 3, pp. 211–230, 2003.

[16]  V. Franzoni and A. Milani, "Structural and semantic proximity in information networks," *Lect. Notes Comput. Sci. (including Subser. Lect. Notes Artif. Intell. Lect. Notes Bioinformatics)*, vol. 10404, pp. 651–666, 2017.

[17]  V. Franzoni and A. Milani, "A pheromone-like model for semantic context extraction from collaborative networks," in *Proceedings - 2015 IEEE/WIC/ACM International Joint Conference on Web Intelligence and Intelligent Agent Technology, WI-IAT 2015*, 2016, vol. 2016–Janua, pp. 540–547.

[18]  V. Franzoni, A. Milani, S. Pallottelli, C. H. C. Leung, and Y. Li, "Context-based image semantic similarity," in *2015 12th International Conference on Fuzzy Systems and Knowledge Discovery, FSKD 2015*, 2016, pp. 1280–1284.

[19]  S. Pallottelli, V. Franzoni, and A. Milani, "Multi-path traces in semantic graphs for latent knowledge elicitation," in *Proceedings - International Conference on Natural Computation*, 2016, vol. 2016–Janua, pp. 281–288.

[20]  V. Franzoni and A. Milani, "A semantic comparison of clustering algorithms for the evaluation of web-based similarity measures," in *Lecture Notes in Computer Science (including subseries Lecture Notes in Artificial Intelligence and Lecture Notes in Bioinformatics)*, 2016, vol. 9790, pp. 438–452.

[21]  V. Franzoni, M. Mencacci, P. Mengoni, and A. Milani, "Semantic heuristic search in collaborative networks: Measures and contexts," in *Proceedings - 2014 IEEE/WIC/ACM International Joint Conference on Web Intelligence and Intelligent Agent Technology - Workshops, WI-IAT 2014*, 2014, vol. 1, pp. 187–217.

[22]  M. E. J. Newman, "Clustering and preferential attachment in growing networks," *Phys. Rev. E*, vol. 64, no. 2, p. 025102, 2001.

[23]  L. Katz, "A new status index derived from sociometric analysis," *Psychometrika*, vol. 18, no. 1, pp. 39–43, 1953.

[24]  V. Franzoni, A. Chiancone, and A. Milani, "A Multistrain Bacterial Diffusion Model for Link Prediction," *Int. J. Pattern Recognit. Artif. Intell.*, vol. 31, no. 11, 2017.

[25]  A. Chiancone, A. Milani, V. Poggioni, S. Pallottelli, A. Madotto, and V. Franzoni, "A multistrain bacterial model for link prediction andrea chiancone," in *Proceedings - International Conference on Natural Computation*, 2016, vol. 2016–Janua, pp. 1075–1079.

[26]  K. W. Church and P. Hanks, "Word association norms, mutual information, and lexicography," in *Proceedings of the 27th annual meeting on Association for Computational Linguistics -*, 1989, pp. 76–83.

[27]  V. Franzoni and A. Milani, "Semantic context extraction from collaborative networks," in *Proceedings of the 2015 IEEE 19th International Conference on Computer Supported Cooperative Work in Design, CSCWD 2015*, 2015, pp. 131–136.

[28]  V. Franzoni, A. Milani, and G. Biondi, "SEMO: A semantic model for emotion recognition in web objects," in *Proceedings - 2017 IEEE/WIC/ACM International Conference on Web Intelligence, WI 2017*, 2017, pp. 953–958.